\documentclass[a4paper,fleqn,11pt]{article}
\usepackage{epsfig}
\usepackage[latin1]{inputenc}
\usepackage{float}
\usepackage{graphicx}
\usepackage{cancel}
\usepackage{mathrsfs}
\usepackage{amssymb}

\author{D'Eramo Francesco}
\title{Dark matter and Higgs boson physics}
\begin{document}
\begin{center}
\textbf{\huge{Dark matter and Higgs boson physics}}
\newline
\newline
\Large{Francesco D'Eramo}
\newline
\textit{\small{Scuola Normale Superiore, Piazza dei Cavalieri 7,
I-56126 Pisa, Italy}}
\end{center}
\vspace{0.2cm}
\begin{abstract}
A vector-like colorless fermion doublet and a singlet added to the
Standard Model allow a consistent interpretation of dark matter in
terms of the lightest neutral particle, as they may help in
obtaining successful gauge coupling unification. We analyze in
detail the mass range of the lightest neutral particle below the $W$
mass, i.e. in a range of the parameters where the physics of the
Standard Model Higgs boson may be substantially affected either
directly or indirectly.

\end{abstract}

\section{Introduction}
The Standard Model (SM) of the electroweak interactions is more than
30 years old and it has been able to reproduce with great precision
the many experimental results obtained until now. In particular at
LEP the theory was tested at the per mille level without finding any
discrepancy with the theoretical predictions. However, in spite of
this extraordinary success, we are convinced of new physics beyond
the SM, since there are problems where the SM does not provide an
adequate solution. One of these, supported by observations, is the
lack of a dark matter candidate.

The most direct and impressive evidence of the existence of dark
matter are surely the flat rotation curves of spiral galaxies. Other
evidences for dark matter were found at different scales, from
galactic scales (several kiloparsecs) and clusters of galaxies
(Megaparsecs) to global scales of hundreds of Megaparsecs
\cite{Bertone:2004pz}. The total matter density can be inferred from
the measurements of the power spectrum of the fluctuations in the
Cosmic Microwave Background (CMB). The recent measurements of the
Wilkinson Microwave Anisotropy Probe (WMAP) \cite{Spergel:2006hy}
have shown that the total matter abundance in the universe is
$\Omega_{m}h^{2}=0.1277^{+0.0080}_{-0.0079}$; these measurements
have provided also the baryon abundance which is
$\Omega_{b}h^{2}=0.02229\pm 0.00073$. We conclude that all the
matter in the universe cannot be baryonic and that the dark matter
abundance is $\Omega_{DM}h^{2}=0.1054^{+0.0080}_{-0.0079}$. The fact
that a significant part of dark matter must be non baryonic was
known before WMAP measurements. Indeed an estimate of
$\Omega_{m}h^{2}$ was already available \cite{Bergstrom:2000pn}, and
the value of $\Omega_{b}h^{2}$ was inferred from primordial
nucleosynthesis \cite{Olive:1999ij}; the difference seemed notable
also at that time. An additional evidence for the non baryonic
nature of the dark matter is given by structure formations: in a
universe with only baryons the primordial density perturbations have
not had enough time to grow and generate the galaxies observed today
in the sky. These observations, however, do not tell us anything
about the particle nature of dark matter. Then the question is about
the nature, the origin and the composition of this important
component of our universe, since dark matter does not find an
explanation in the framework of the Standard Model of particle
physics.

Particle physics provides us with a large number of dark matter
candidates, which appear naturally in various frameworks for reasons
completely independent from the dark matter problem, and certainly
not invented for the sole purpose of explaining the presence of dark
matter in our universe. Among these candidates an important
distinction is between particles created thermally or not thermally
in the early universe. For thermal relics another important
distinction is about how they decoupled from the primordial soup, in
particular if they were relativistic (hot dark matter) or
nonrelativistic (cold dark matter). Arguments from large structures
make us believe that a large, and presumably dominant, fraction of
dark matter is made of cold relics. A well motivated class of cold
dark matter particles are the so called Weakly Interacting Massive
Particles (WIMPs), which have mass between $10\;GeV$ and a few $TeV$
and interact only through weak and gravitational interactions,
because the limits on charged relics are very stringent
\cite{pdgcaric}.

Another missing opportunity for the Standard Model is that gauge
couplings do not quite unify at high energy; a possible solution is
to add weakly interacting particles to change the running, in order
to make unification work better.

In this work we discuss a model that has both a cold dark matter
candidate and can improve considerably over the Standard Model in
the direction of successful gauge coupling unification. We introduce
new matter with respect to the Standard Model alone, and we restrict
ourselves to the case in which the added particles are fermions.
Adding just a vector doublet allows remarkable improvements for
unification; this model, furthermore, is highly constrained since it
contains only one new parameter, the Dirac mass for the degenerate
doublets, whose neutral components are the dark matter candidates.
Such model, however, is ruled out by direct detection experiments:
the vector-like vertex with the $Z$ boson for the neutral particles
remains unsuppressed, giving a spin-independent cross section that
is $2$-$3$ orders of magnitude above current limits
\cite{Cirelli:2005uq}. This drawback can be solved by including a
fermion singlet, with Yukawa couplings with the doublets and the
Higgs boson. Doing this we generate a mixing between doublets and
singlet, so that the neutral particles become Majorana fermions
which have suppressed vector-like couplings with the $Z$ boson. We
assume a parity symmetry that acts only on the new fields. This
imposes that they do not couple to ordinary matter. It also implies
that the lightest particle is stable and, if neutral, it constitutes
a good dark matter candidate. This model has been introduced in
reference \cite{Mahbubani:2005pt} where a detailed dark matter
analysis for high values of the relic particle mass can be found. In
reference \cite{Mahbubani:2005pt} it is also shown how the gauge
coupling unification at high energy can be achieved and a rate for
the proton decay is predicted that could be tested in the future.

In this work we focus on the region of parameter space where the
mass of the lightest neutral particle (LNP) is smaller than the $W$
boson mass $m_{W}\approx80\,GeV$. The analysis for higher mass was
already done, as said above, but the main reason for doing so is
that well above the $WW$ production threshold, in order to account
for the entire dark matter abundance observed, the mass $M$ of
charged components of the doublets is quite high.  An important fact
is that, for relatively low values of the LNP mass, the effects on
Higgs boson physics are significant, both direct and indirect. On
the one hand there are new available decay channels for the Higgs
boson, and the decays into neutral particles may dominate the total
width. On the other hand the new particles contribute to electroweak
observables, so that they may change the indirect upper limit on the
Higgs mass and improve the naturalness of the Higgs potential
\cite{Barbieri:2006dq}. There are thus reasons to give special attention to this
region of parameter space.

The structure is the following: in section \ref{model} we present
the model with its spectrum, in section \ref{DM} we compute the
relic abundance of the dark matter candidate, in section
\ref{DirDet} we discuss direct detection, in section \ref{Higgs} the
effects on Higgs boson physics. Finally in section \ref{EDMch} we
consider a possible $CP$ violating phase giving rise to an electron
electric dipole moment. Conclusions are given in section
\ref{Conclusion}.

\section{The model}\label{model}
The model consists of the following extension of the Standard Model
\begin{equation}
\mathcal{L} = \mathcal{L}_{SM}+\Delta \mathcal{L}
\end{equation}
where we add to the Standard Model lagrangian the following
renormalizable lagrangian (other than the kinetic terms for the
various new fields)
\begin{equation}
\Delta \mathcal{L} = \lambda F H S + \lambda^{c} F_{c} H^{\dag} S +
M F F_{c} + \mu S^{2} + h.c.
\end{equation}
The doublets $F_{c}$ and $F$ have respectively hypercharge $\pm
1/2$, $S$ is a singlet and $H$ is the Standard Model Higgs doublet.
We introduce the symmetry
\begin{equation}
F, \, F_{c}, \, S \rightarrow -F, \, -F_{c}, \, -S \label{symmetry}
\end{equation}
with all other fields invariant. This imposes that the new fields do
not couple to ordinary matter. We suppose the parameters
$\left(\lambda,\,\lambda^{c},\,M,\,\mu\right)$ to be real (in
section \ref{EDMch} we will consider the effects of introducing a
phase). The physical fields are chosen as follows
\begin{equation}
F_{c}=\left( \begin{array}{c}
F^{+}\\
F^{0}_{c}
\end{array} \right)
\;\;\;\;\;\;\;\;\;\;\;\; F=\left( \begin{array}{c}
F^{0}\\
F^{-}
\end{array} \right)
\;\;\;\;\;\;\;\;\;\;\;\; H=\left( \begin{array}{c}
\phi^{+}\\
v+\frac{h+i \chi}{\sqrt{2}}
\end{array} \right)
\label{campi}
\end{equation}
The components of $F_{c}$ and $F$ are left-handed Weyl fields. The
Goldstones $\phi^{+}$ and $\chi$ can be put to zero by choosing the
unitary gauge.

In the charged sector there is a simple Dirac term of mass $M$,
hence we define the Dirac spinor
$\psi=F_{c}^{+}+\left(F^{-}\right)^{c}$

In the neutral sector we define the fields $N_{i}$ as
\begin{equation}
N_{1}=\frac{1}{\sqrt{2}}\left(F^{0}_{c}-F^{0} \right)
\;\;\;\;\;\;\;\;\;\;\;\;
N_{2}=\frac{1}{\sqrt{2}}\left(F^{0}_{c}+F^{0} \right)
\;\;\;\;\;\;\;\;\;\;\;\;N_{3}=S \label{cambiobase}
\end{equation}
so that the mass matrix takes the form
\begin{equation}
M_{N}=\left(
\begin{array}{ccc}
M & 0 & -\sqrt{2}\beta v \\
0 & -M & -\sqrt{2}\alpha v\\
-\sqrt{2}\beta v & -\sqrt{2}\alpha v & -2 \mu
\end{array} \right)
\label{massmatrixN}
\end{equation}
where the Yukawa couplings have been replaced by the parameters
\begin{equation}
\alpha=\frac{\lambda^{c}+\lambda}{2} \;\;\;\;\;\;\;\;\;\;\;\;
\beta=\frac{\lambda^{c}-\lambda}{2} \label{alphaandbeta}
\end{equation}
We have to find now eigenvalues and eigenvectors of this matrix; in
the general case this task cannot be accomplished analytically and
therefore we will diagonalize the mass matrix numerically. Let
$m_{i}$ be the eigenvalues and let $V$ be the matrix that performs
the diagonalization. We define $\chi_{i}$ as the eigenvector
corresponding to $m_{i}$, i.e.
\begin{equation}
N_{i}=V_{ij}\chi_{j} \;\;\;\;\;\;\;\;\;\;\;\;\;\; V^{t}M_{N}V
=diag\left(m_{1}, m_{2}, m_{3}\right) \label{diagonalizzazionemassa}
\end{equation}
We identify the lightest neutral particle (LNP) with the index $l$,
then $\chi_{l}$ is the field of the LNP\footnote{from now on the
index $l$ for $\chi_{l}$ indicates \emph{lightest}, and it must not
be confused with \emph{left}}.

\section{Dark matter analysis}\label{DM}
In this section we compute the thermal relic abundance of the LNP
using the standard formalism \cite{1990eaun.book.....K}. Before
proceeding we should justify why we can use it, because there are
situations in which this method fails \cite{Griest:1990kh}. We have
checked that in the parameter region of interest to us the masses of
the other two neutral and of the charged particles are far higher
than the LNP mass itself, so we can neglect coannihilations. The
standard method is also not valid when the relic particle lies near
a mass threshold since the LNP particles are Boltzmann distributed.
Given our LNP mass range the only threshold present is that for $WW$
production. For $m_{l}\geq 75\,GeV$ the $WW$ process suppresses the
LNP relic abundance to an unacceptable level, whereas for $m_{l} <
75\,GeV$ it can be safely neglected.

The evolution of the LNP number density $n_{l}$ is governed by the
Boltzmann equation
\begin{equation}
\frac{dn_{l}}{dt}+3Hn_{l}=-<\sigma
v_{rel}>\left[n_{l}^{2}-\left(n_{l}^{eq}\right)^{2}\right]
\label{BoltzEquat}
\end{equation}
where $H$ is the Hubble parameter, $n_{l}^{eq}$ is the LNP
equilibrium number density, $v_{rel}$ is the relative velocity and
$<\sigma v_{rel}>$ is the thermal average of the annihilation cross
section. The relevant temperatures are of order $m_{l}/25$, so the
Boltzmann equilibrium distribution is well justified. The Boltzmann
equation can be solved approximately. First we introduce the
variable $x \equiv m_{l}/T$. Secondly we parameterize the
temperature dependence of the annihilation cross section as
\begin{equation}
<\sigma v_{rel}>=\sigma_{0}\,x^{-n}
\end{equation}
where $n=0$ corresponds to s-wave annihilation, $n=1$ to p-wave
annihilation, etc. At early times $n_{l}$ is accurately approximated
by $n_{l}^{eq}$, but as the temperature drops below the mass
$m_{l}$, $n_{l}^{eq}$ drops exponentially until a point denominated
``freeze out'' is reached where the reaction rate is not fast enough
to maintain equilibrium. From this point on, the $n_{l}^{eq}$ term
in equation (\ref{BoltzEquat}) can be neglected and the remaining
equation is easily integrated. Thus the solution of
(\ref{BoltzEquat}) is given by solving in two regimes and matching
those solutions at the freeze out. The value of the freeze out point
$x_{f}$ is obtained by imposing the equality between the interaction
rate $\Gamma = n_{l}\sigma v_{rel}$ and the expansion rate $H$, and
it is given by the numerical solution of the following equation
\begin{equation}
x_{f} + \left(n+\frac{1}{2}\right)\ln x_{f} =
\ln\left[0.038\left(n+1\right)\,\left(g/g_{*}^{1/2}\right)\,
 m_{Pl}\, m_{l}\,\sigma_{0}\right]
\end{equation}
where the Planck mass is $m_{Pl}=1.22\times10^{19}\,GeV$ and $g_{*}$
is the number of effectively relativistic degrees of freedom at the
time of freeze out. The present mass density of the relic particles
is expressed as
\begin{equation}
\Omega_{l}
h^{2}=\left(n+1\right)\frac{x_{f}^{n+1}}{g_{*}^{1/2}}\frac{0.034\;pb}{\sigma_{0}}
\end{equation}

In our model we are dealing with a cold relic, therefore in the
early universe, just before the decoupling, thermal equilibrium is
maintained via LNP annihilations into fermions. There are two
possible processes: $Z$ boson exchange and Higgs boson exchange,
both p-wave. Cross sections for these processes and their thermal
averages are given in appendix \ref{app-cross}. In the following
discussion we fix the values of Yukawa couplings and analyze the
model as function of $\left(\mu,M\right)$ for each case. The limit
of small Yukawa couplings $\lambda$, $\lambda^{c}$ is not
interesting, since in this case the LNP coincides approximately with
the singlet, and the only way to produce all the dark matter
observed is with the LNP mass near the $Z$ pole or the Higgs pole.
If $\lambda^{c}=\lambda$ the model possesses a $SU(2)_{L}\times
SU(2)_{R}$ symmetry broken to $SU(2)_{V}$ by the Higgs vacuum
expectation value, and the coupling with the $Z$ boson is
suppressed. Also in this case the only way to produce all the dark
matter is near the Higgs pole.

We consider for the complete analysis two limiting cases: almost
equal Yukawa couplings (symmetric case, or more properly nearly
symmetric) and when one of them is vanishingly small (asymmetric
case). To be consistent with negative searches from LEP we assume
$m_{l}\geq45\,GeV$ and $M\geq100\,GeV$. The cases which we discuss
are (the reason for doing so is explained in section \ref{Higgs})
\begin{itemize}
\item Symmetric case: $\alpha=1.0$ and $\beta=0.1$
\item Asymmetric case I: $\alpha=0.5$ and $\beta=0.5$
\item Asymmetric case II: $\alpha=0.65$ and $\beta=0.65$
\end{itemize}

Before proceeding we must say something about the Higgs boson mass,
since the annihilation cross section for Higgs exchange depends on
it and we have to choose its value carefully. We will see in section
\ref{Higgs} that in the symmetric case the corrections to the
electroweak parameter $T$ are negligible, then the indirect upper
limit on the Higgs mass valid in the Standard Model ($m_{h} \lesssim
166\,GeV$ at $95\%$ CL \cite{Alcaraz:2006mx}) remains unchanged. On
the contrary, in the asymmetric cases $T$ is strongly affected by
the new particles, so the upper limit is raised. We choose the
reference values
\begin{itemize}
\item Symmetric case: $m_{h}=120\,GeV$
\item Asymmetric case I and II: $m_{h}=300\,GeV$
\end{itemize}

\begin{figure}
\centering
\includegraphics[scale=0.82]{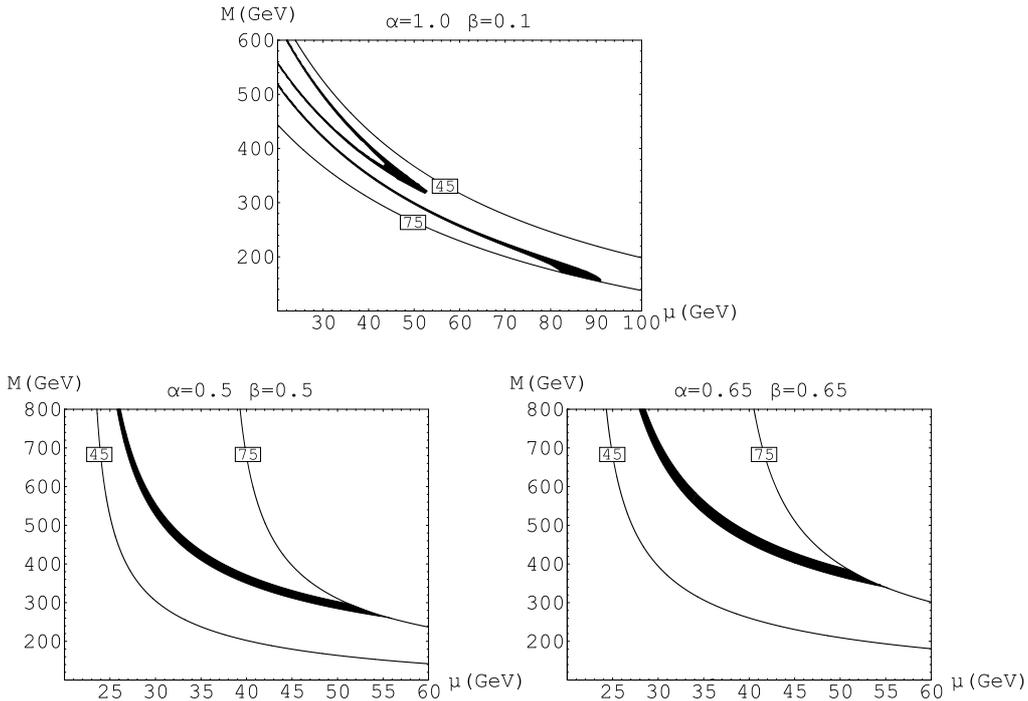}
\caption{LNP relic abundance. Contours for $m_{l}=45,\,75\, GeV$ are
denoted by the solid lines, shaded regions correspond to
$0.089\leq\Omega_{l} h^{2}\leq 0.122$ (WMAP 95\%CL region).}
\label{DMplot}
\end{figure}

We plot our results in the $\left(\mu,M\right)$ plane in figure
\ref{DMplot}. In the symmetric case only the relative sign of $\mu$
and $M$ is physical, and our convention is $M>0$. In the asymmetric
case both signs are unphysical, then we choose $\mu>0$ and $M>0$. We
identify the parameter space region for which $45\,GeV \leq m_{l}
\leq 75\,GeV$ and inside it we shade the area for which
$0.089\leq\Omega_{l} h^{2}\leq 0.122$ (corresponding to the 95\%CL
region from WMAP \cite{Spergel:2006hy}). In all the cases the dark matter
abundance can be accounted for by our LNP.

Another check must be done: neutral particles could have been
produced at LEP2
\begin{equation}
e^{+}e^{-} \rightarrow \chi_{l}\, \chi_{nl} \label{lnlatlep2}
\end{equation}
where the index $nl$ stands for ``next to lightest''. Given the
assumed symmetry (\ref{symmetry}) the only allowed decay for
$\chi_{nl}$  is
\begin{equation}
\chi_{nl} \rightarrow \chi_{l}\,f\,\overline{f} \label{lnlff}
\end{equation}
where $f$ indicates a generic fermion and $\overline{f}$ the
corresponding antifermion. Since no such event was seen this may
constrain the model. We have checked that it has not been
kinematically allowed at LEP2, since the ``next to lightest''
particle mass is always above $200\,GeV$ in the parameter space
region of interest.

\section{Direct detection}\label{DirDet}
Dark matter particles of the Milky Way might be detectable as they
pass through detectors in laboratories on Earth. The very low cross
section of WIMPs on ordinary material makes these interactions quite
rare, but recent experiments have made progress. The direct
detection experiments can measure and distinguish from background
the tiny energy deposited by elastic scattering of a WIMP off a
target nucleus. The current experimental results set limits on
WIMP-nucleon cross sections, and we compare LNP-nucleon cross
sections given in appendix \ref{DirDetApp} with these limits. Dark
matter particles in the Milky Way halo have presumably a mean speed
$<v>\,\simeq\, 300\, Km\, s^{-1} = 10^{-3} c$, therefore the process
can be treated in the nonrelativistic limit.

The nucleon coupling of a slow-moving Majorana fermion is
characterized by two terms: spin-dependent (axial-vector) and
spin-independent (scalar). We consider these two contributions
separately.

The spin-dependent cross section for LNP-nucleus elastic scattering
is given by (\ref{formulaspinDEP}). For a proton target
$\Lambda^{2}J\left(J+1\right)\simeq 1$ and the cross section is
\begin{equation}
\sigma_{Z}\left(LN\rightarrow
LN\right)=3.5\left(V_{1l}V_{2l}\right)^{2}\times 10^{-1}\;pb
\label{crossspindepNUM}
\end{equation}
(for the definition of $V$ see (\ref{diagonalizzazionemassa})). The
cross section (\ref{crossspindepNUM}) for the three cases discussed
above is always $2-3$ orders of magnitude below current limits
\cite{Akerib:2005za}.

\begin{figure}
\centering
\includegraphics[scale=1]{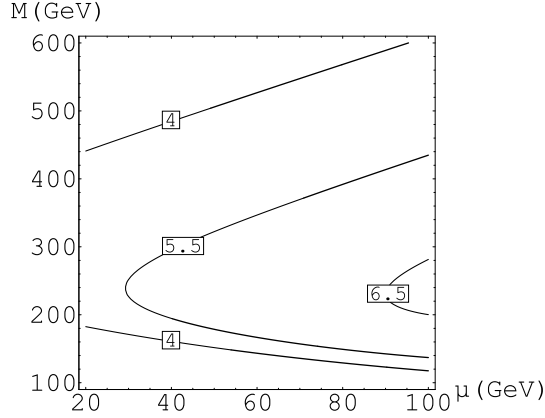}
\caption{Spin-independent cross section: symmetric case (units
$10^{-7} \,  pb$)} \label{H1}
\end{figure}

The spin-independent cross section is given by
(\ref{formulaspinINDEP}). It depends sensibly on the Higgs mass, so
it is different between symmetric and asymmetric cases. For
scattering from a proton
\begin{eqnarray}
\sigma_{h}\left(LN\rightarrow LN\right) &&= 2.75\,\xi^{2}\times
10^{-6}\;pb\;\left(\frac{120\, GeV}{m_{h}}\right)^{4} {}\nonumber\\
& & {}=7.04\,\xi^{2}\times 10^{-8}\;pb\;\left(\frac{300\,
GeV}{m_{h}}\right)^{4} \label{crossspinindepNUM}
\end{eqnarray}
where $\xi = V_{3l} \left(\alpha V_{2l} + \beta V_{1l} \right)$. For
the reason explained in section \ref{Higgs}, we take the former
reference value in the symmetric case and the latter in the
asymmetric cases. The cross section (\ref{crossspinindepNUM}) for
the symmetric case is plotted in figure \ref{H1} in units of
$10^{-7} \, pb$. It is about one order of magnitude above the
experimental limits \cite{Angle:2007uj}. In the asymmetric cases,
instead, spin-independent cross section is always $1-2$ orders of
magnitude below current limits,  but within the sensitivity of
experiments currently under study \cite{Brink:2005ej}.

\section{Higgs boson physics}\label{Higgs}
In this section we analyze the effects on Higgs boson physics
induced by the new particles. In the first subsection we compute the
contributions to the electroweak observables from their virtual
exchanges and we will see that the upper limit on the Higgs mass is
significantly affected. In the second subsection we analyze the new
available decay channels to Higgs boson decays and compute the
relevant branching ratios.

\subsection{ElectroWeak Precision analysis}
The interaction lagrangian of the new particles with the gauge
bosons is
\begin{eqnarray}
\Delta \mathcal{L}|_{int} & = & -V_{1i}\frac{g}{2}\, W_{\mu}^{+}
\overline{\psi}\,\gamma^{\mu}\,\chi_{i}+\, h.c. + {}\nonumber\\ & &
{}V_{2i}\frac{g}{2}\, W_{\mu}^{+}\overline{\psi}\,\gamma^{\mu}\gamma^{5}\,\chi_{i} \, + \, h.c. + {}\nonumber\\
& & {}
\frac{g}{2}W_{\mu}^{3}\left[\overline{\psi}\,\gamma^{\mu}\,\psi+
\frac{1}{2}\left(V_{1i}V_{2j}+V_{2i}V_{1j}\right)
\overline{\chi_{i}}\,\gamma^{\mu}\gamma^{5}\,\chi_{j} \right]+
{}\nonumber\\ & & {}
\frac{g^{'}}{2}B_{\mu}\left[\overline{\psi}\,\gamma^{\mu}\,\psi-
\frac{1}{2}\left(V_{1i}V_{2j}+V_{2i}V_{1j}\right)
\overline{\chi_{i}}\,\gamma^{\mu}\gamma^{5}\,\chi_{j} \right]
\label{gaugebosonsinteraction}
\end{eqnarray}
where sums under repeated indices are understood. The new particles
contributions to $T$ and $S$ are respectively
\begin{eqnarray}
T & = & \sum_{i=1}^{3}\left[
\left(V_{1i}\right)^{2}\widetilde{A}\left(M,m_{i}\right)+
\left(V_{2i}\right)^{2}\widetilde{A}\left(M,-m_{i}\right)\right]
{}\nonumber\\ & & {} -\frac{1}{2}\sum_{i,j=1}^{3}
\left(V_{1i}V_{2j}+V_{2i}V_{1j}\right)^{2}\widetilde{A}\left(m_{i},-m_{j}\right)
\end{eqnarray}
\begin{equation}
S =   \frac{1}{2}\sum_{i,j=1}^{3}
\left(V_{1i}V_{2j}+V_{2i}V_{1j}\right)^{2}\widetilde{F}\left(m_{i},-m_{j}\right)-\widetilde{F}\left(\mu,\mu\right)
\end{equation}
The functions $\widetilde{F}$ and $\widetilde{A}$ are defined in
appendix \ref{EWPTappendix}.

\begin{figure}
\centering
\includegraphics[scale=0.52]{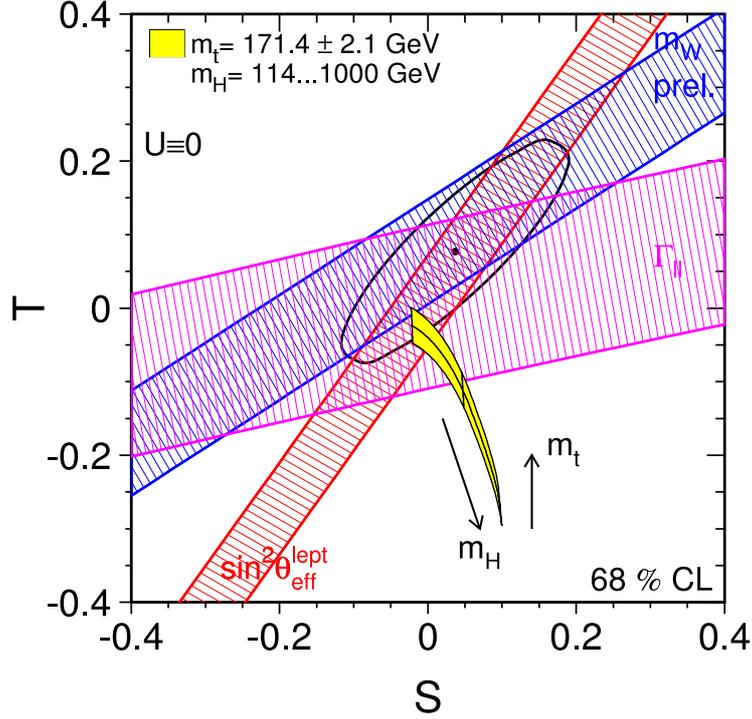}
\caption{Region of the $\left(S,T\right)$ plane allowed by EWPT at
$68\%$CL and dependence of $S$ and $T$ on the Higgs mass. The thin
black line marks $m_{h}=400\,GeV$. From \cite{stcon}.}
\label{STcontours}
\end{figure}

\begin{figure}
\centering
\includegraphics[scale=0.8]{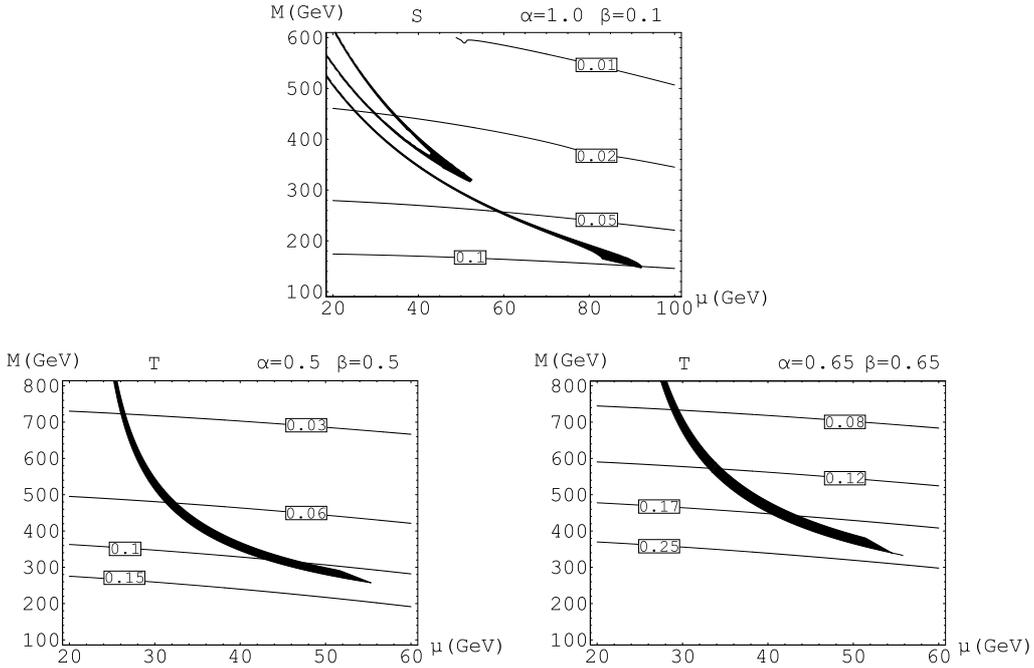}
\caption{S in the symmetric case and T in the asymmetric cases. The
shaded regions are such that $0.089\leq\Omega h^{2}\leq 0.122$.}
\label{EWPT}
\end{figure}

We have now all the ingredients to perform the analysis. We have
verified that in the symmetric case the contribution to $T$ is
negligible, as required by the custodial symmetry mentioned in
section \ref{DM}, whereas in the asymmetric cases $S$ is not
significantly affected. The experimental contours in the
$\left(S,T\right)$ plane are shown in figure \ref{STcontours} and
our results for the significant cases are shown in figure
\ref{EWPT}.

In the symmetric case $\Delta T$ is irrelevant and $\Delta S$ is
inside the experimental ellipse for almost all the region that
provides the entire dark matter abundance; if we raise the value of
Yukawa couplings then $\Delta S$ goes rapidly outside the ellipse,
so we restrict ourselves just to this symmetric case and we do not
consider higher values of $\alpha$. Looking at figure
\ref{STcontours} one can immediately see how an heavy Higgs can be
allowed by ElectroWeak Precision Tests (EWPT): the only thing that
we need is new physics producing a positive $\Delta T$ and a not too
large $\Delta S$. To raise the Higgs mass up to $500\,GeV$ the
needed compensation is $\Delta T\approx0.2$ \cite{Barbieri:2006dq}.
The asymmetric case is perfectly suited to this purpose: it gives
unimportant $\Delta S$ and a positive $\Delta T$ as desired. We
first studied the case of Yukawa coupling $\lambda^{c}$ equal to
$1$, then we raised its value until we reached $\Delta T\approx0.2$,
and this corresponds to $\lambda^{c}=1.3$ or equivalently
$\alpha=0.65$ . All the values of $\lambda$ and $\lambda^{c}$ that
we consider are consistent with a Landau pole for the Yukawa
coupling above the unification scale. When the Higgs boson mass is
raised, however, the Higgs quartic coupling is very likely to have a
Landau pole below the unification scale.

\subsection{Higgs boson decays}
Another important effect on the Higgs boson physics is the increase
of its total width. In the parameter space region of interest to us
the only new available decay channel is  $h\rightarrow
\chi_{l}\chi_{l}$, since the decays into other new particles are
kinematically forbidden. The partial width for such decay results in
\begin{equation}
\Gamma_{\chi\chi}=\xi^{2}\frac{m_{h}}{2\pi}\left(1-\frac{4m^{2}}{m_{h}^{2}}\right)^{\frac{3}{2}}
\label{partialwidth}
\end{equation}
where the parameter $\xi$ is defined in (\ref{Higgscoupling}).  The
Higgs total width predicted by the Standard Model $\Gamma_{h}^{SM}$
is known as a function of $m_{h}$ \cite{Djouadi:2005gi}. We consider four
values for the Higgs mass, and the correspondent SM width are
reported here
\begin{center}
\begin{tabular}{|c|c|}
\hline
$m_{h}\left(GeV\right)$ & $\Gamma_{h}^{SM}\left(GeV\right)$ \\
\hline
$120$ &  $3.65 \times 10^{-3}$\\
\hline
$150$ & $1.67 \times 10^{-2}$\\
\hline
$200$ & $1.425$\\
\hline
$300$ & $8.50$\\
\hline
\end{tabular}
\end{center}
The partial width $\Gamma_{\chi\chi}$ for decay into two LNPs is
given by (\ref{partialwidth}), and thus we can compute the branching
ratio
\begin{equation}
BR\left(h\rightarrow
\chi\chi\right)=\frac{\Gamma_{\chi\chi}}{\Gamma_{h}^{SM}+\Gamma_{\chi\chi}}
\end{equation}

For $m_{h}=120\,GeV$ we compute the branching ratio in the symmetric
case, whereas for higher masses we make the calculation for both the
asymmetric cases. As seen from figure \ref{DMplot} in both the
asymmetric cases the only free parameter is $M$, since if we impose
$\Omega_{l} h^{2}=0.105$ the value of $\mu$ is automatically fixed.
This is not true for the symmetric case, where for each $M$ there
are up to three values of $\mu$. The branching ratios are plotted as
a function of $M$. In the asymmetric cases it is the only free
parameter. In the symmetric case we consider the line of figure
\ref{DMplot} corresponding to the lower value of the LNP mass. For
the line corresponding to higher value of the LNP mass the decay is
kinematically forbidden. For the line in the middle even small pole
effects might modify the branching ratios considerably, because we
are in a region where the phase space is nearly saturated.

\begin{figure}
\centering
\includegraphics[scale=0.8]{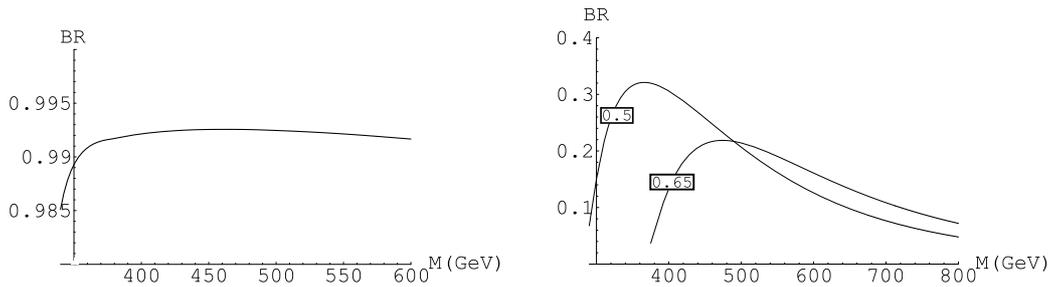}
\caption{Branching ratio as a function of $M$ for $m_{h}=120\,GeV$
(left) and for $m_{h}=150\,GeV$ (right). Labels indicate the value
of $\alpha$.} \label{BR}
\end{figure}

The branching ratios for $m_{h}=120\,GeV$ and for $m_{h}=150\,GeV$
are plotted in figure \ref{BR}. In the simmetric case this channel
dominates the total width. For higher values of the Higgs boson mass
the branching ratios decrease, as a consequence that the Standard
Model width increases faster than the partial width into two LNPs.
For $m_{h}=200\,GeV$ they are always below the $4\%$, while for
higher values of the Higgs mass they are even smaller.

\section{Electric dipole moment}\label{EDMch}
We have taken the parameters
$\left(\lambda,\,\lambda^{c},\,M,\,\mu\right)$ to be real until now.
We now explore the possibility of a $CP$ violating phase. This phase
could be present only in the symmetric case, since if one of the
Yukawa couplings vanishes (as in the asymmetric case) all the
parameters can be made real by a fields redefinition. In the general
case we can redefine fields so that
$\left(\lambda,\,\lambda^{c},\,\mu\right)$ are real, leaving a
residual phase on the parameter $M$. The mass matrix $M_{N}$ found
in (\ref{massmatrixN}) becomes
\begin{equation}
M_{N}=\left(
\begin{array}{ccc}
M e^{i\theta} & 0 & -\sqrt{2}\beta v \\
0 & -M e^{i\theta} & -\sqrt{2}\alpha v\\
-\sqrt{2}\beta v & -\sqrt{2}\alpha v & -2 \mu
\end{array} \right)
\end{equation}

\begin{figure}
\centering
\includegraphics[scale=0.68]{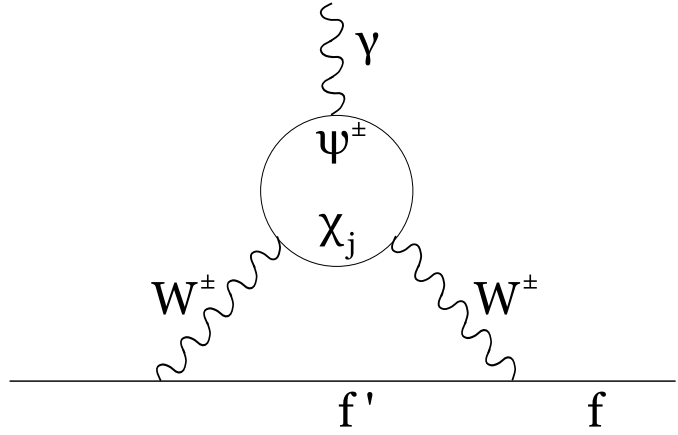}
\caption{Two loop contribution to the electric dipole moment of a
fermion f} \label{2loopsEDM}
\end{figure}

The phase $\theta$ induces an electron electric dipole moment (EDM)
at two loops, the dominant diagram responsible for it is generated
by charged and neutral particles and is shown in figure
\ref{2loopsEDM} \cite{Mahbubani:2005pt}.
The induced EDM moment is given by
\begin{equation}
\frac{d^{W}_{f}}{e}=\pm\frac{\alpha^{2}m_{f}}{8\pi^{2}s^{4}_{W}m_{W}^{2}}
\sum^{3}_{i=1}\frac{m_{\chi_{i}}M}{m_{W}^{2}}Im\left(O^{L}_{i}O^{R*}_{i}\right)\mathcal{G}\left(r_{i}^{0},r^{\pm}\right)
\label{edmformula}
\end{equation}
where
\begin{eqnarray}
\mathcal{G}\left(r_{i}^{0},r^{\pm}\right) & = &
\int_{0}^{+\infty}dz\int_{0}^{1}\frac{d\gamma}{\gamma}\int_{0}^{1}dy
\frac{yz\left(y+z/2\right)}{\left(z+y\right)^{3}\left(z+K_{i}\right)}=
{}\nonumber\\ & & {}
\int_{0}^{1}\frac{d\gamma}{\gamma}\int_{0}^{1}dy\,y
\left[\frac{\left(y-3K_{i}\right)y+2\left(K_{i}+y\right)y}{4y\left(K_{i}-y\right)^{2}}+
\frac{K_{i}\left(K_{i}-2y\right)}{2\left(K_{i}-2y\right)^{3}}\ln\frac{K_{i}}{y}\right]{}\nonumber\\
& & {}
\end{eqnarray}
and
\begin{equation}
\begin{array}{l}
K_{i}=\frac{r^{0}_{i}}{1-\gamma}+\frac{r^{\pm}}{\gamma}\,; \;\;\;\;
r^{\pm}\equiv\frac{M^{2}}{m_{W}^{2}}\,;
\;\;\;\;r^{0}_{i}\equiv\frac{m_{\chi_{i}}^{2}}{m_{W}^{2}}\\
O^{R}_{i}=\sqrt{2}V_{2i}^{*} \exp\left(-i\theta\right)\,; \;\;\;\;
O^{L}_{i}=-N_{3i}
\end{array}
\end{equation}
The matrix $V$ diagonalizes the mass matrix and is such that
$V^{T}M_{N}V=diag\left(m_{1},m_{2},m_{3}\right)$ with real and
positive diagonal elements. The sign on the right-hand side of
equation (\ref{edmformula}) corresponds to the fermion $f$ with weak
isospin $\pm1/2$ and $f^{'}$ is its electroweak partner.

\begin{figure}
\centering
\includegraphics[scale=0.8]{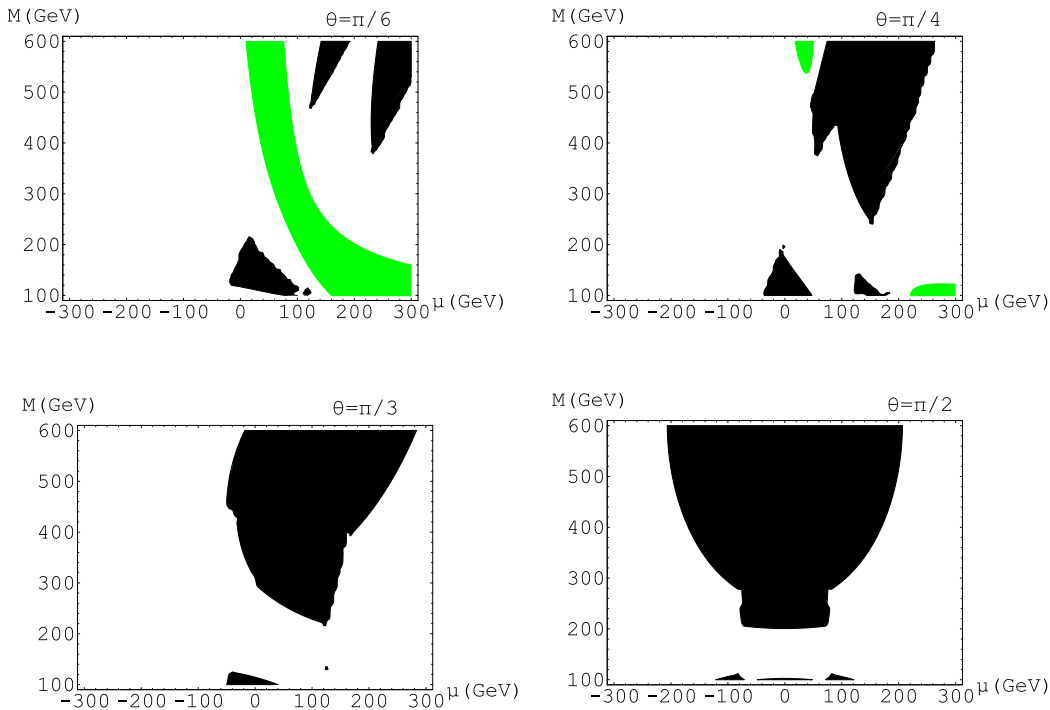}
\caption{Induced electron EDM for $\theta=\frac{\pi}{6},
\frac{\pi}{4},\frac{\pi}{3},\frac{\pi}{2}, $. Green (lightest)
regions are such that $m_{l}\leq75\,GeV$, black shading indicates
regions where the induced EDM is above experimental limit.}
\label{EDM}
\end{figure}

The experimental limit on electron electric dipole moment at
$95\%$CL level is \cite{Regan:2002ta}
\begin{equation}
|d_{e}| < 1.7 \times 10^{-27}\,e\,cm \label{EDMeqlimit}
\end{equation}
We consider four different values of the phase, namely
$\theta=\pi/6,\,\pi/4,\pi/3,\pi/2$, and we shade in the usual
$\left(\mu,M\right)$ plane the regions where the induced EDM is
above such limit. We identify in that plane also the region where
the LNP mass is below $75\,GeV$, since this is the case of our
interest. We also restrict ourselves to charged particle mass $M$
below $600\,GeV$ as in the plot shown in figure \ref{DMplot}. The
plots are shown in figure \ref{EDM}. The first result is that for
$M<600\,GeV$ we can have a LPN mass below $75\,GeV$ only for small
phases, otherwise the imaginary part not present before would
require a cancellation in the mass matrix obtainable only for higher
value of $M$. Regarding the induced EDM there are regions where it
is above the experimental limit, but never inside regions such that
$m_{l}<75\,GeV$. On the contrary, for $m_{l}<75\,GeV$, the induced
EDM is always below the limit (\ref{EDMeqlimit}). It could however
be accessible to next generation experiments
\cite{Semertzidis:2004uu} \cite{Kawall:2004nv}.

\section{Conclusions}\label{Conclusion}
In the last few decades it has been realized that the ordinary
matter which we have been studying until now constitutes only about
$5 \%$ of the total universe energy density. Evidence for
nonluminous gravitating mass abounds on all the scales, from
galactic to global ones of hundreds of Megaparsec. The measurements
of the light element abundances and of the fluctuations in the
Cosmic Microwave Background show that a significant part of the dark
matter must be non baryonic. The Standard Model of particle physics
does not contain such component. Another missing opportunity for the
Standard Model is that gauge coupling unification does not occur at
high energy. In this work we have discussed a minimal extension of
the Standard Model, which can explain all the observed dark matter
abundance and improve the gauge coupling unification, focusing on
the parameter space region for which the LNP mass is below
$m_{W}\approx80\,GeV$. For such region the effects on the Higgs
boson physics are worth of consideration.

We have considered two limiting cases: almost equal Yukawa couplings
(symmetric) and one of them vanishing (asymmetric). In both cases
all the observed dark matter abundance could be explained by the
LNP. We have computed also the full spectra of the model for all the
cases, and they are consistent with negative searches from LEP2. The
spin-independent direct detection cross section is above the current
limits only for the symmetric case. In the asymmetric case it is
well below these limits, as the spin-dependent cross sections for
both cases. However they are all within the sensitivity of
experiments currently under study.

The new particles might have both direct and indirect effects on the
Higgs boson physics. We have analyzed these effects and found that
they are very different for each case. In the symmetric case the
contribution to the electroweak observables is small, but the Higgs
decays in LNP pairs dominate the total width. This might hide the
Higgs boson at the Large Hadron Collider. On the contrary in the
asymmetric case the contribution to the EWPT is important, and the
indirect limit on the Higgs mass valid in the Standard Model can be
raised. Finally we have considered a CP violating phase for the
Dirac mass of the charge particle, giving rise to an electron
electric dipole moment. We have verified that if we keep the LNP
mass below $75\,GeV$ the induced electric dipole moment is always
below the current experimental limit, but perhaps accessible at the
next generation experiments.

\section*{Acknowledgements}
I would like to thank Riccardo Barbieri for his constant presence
and guidance, without which this work would have never been
completed, and for the careful reading of the manuscript. I am
grateful to Vyacheslav S. Rychkov for some precious suggestions
concerning numerical computation and to Alessandro Strumia for
useful discussions concerning direct detection experimental limits.

\appendix
\section{Annihilations cross sections}\label{app-cross}
\subsection{$s$-channel $Z$ exchange}
The LNP has a coupling with the $Z$ boson given by
\begin{equation}
\frac{g}{2 c_{W}}\left(V_{1l}V_{2l}\right)Z_{\mu}
\overline{\psi_{l}}\gamma^{\mu}\gamma^{5}\psi_{l} \label{Zvertex}
\end{equation}
The cross section for the process $\chi_{l}\;\chi_{l}\;\rightarrow
Z^{*}\rightarrow f\;\overline{f}$ for nonrelativistic LNPs and in
the limit of massless final products is
\begin{equation}
\sigma_{Z}\, v_{rel}=\Sigma\left(g_{V}^{2}+g_{A}^{2}\right)
\;\frac{g^{4}\left(V_{1l}V_{2l}\right)^{2} }{24\pi c_{W}^{4}}
\;\frac{m^{2}}{\left(4m^{2}-m_{Z}^{2}\right)^{2}}v_{rel}^{2}
\label{annZ}
\end{equation}
where the sum runs over all the Standard Model fermions except for
the top quark.
\subsection{$s$-channel $h$ exchange}
The coupling between the LNP and the Higgs boson is
\begin{equation}
\xi\overline{\psi_{l}}\psi_{l}\,h
\;\;\;\;\;\;\;\;\;\;\;\;\;\;\;\;\;\;\;\;\;\;\;\;\;\;  \xi \equiv
V_{3l} \left(\alpha V_{2l} + \beta V_{1l}
\right)\label{Higgscoupling}
\end{equation}
The cross section for the process $\chi_{l}\;\chi_{l}\;\rightarrow
h^{*}\rightarrow f\;\overline{f}$, kinematically identical to the
previous one, is
\begin{equation}
\sigma_{h} v_{rel}=
\frac{\xi^{2}}{4\pi}\;\frac{m^{2}m_{b}^{2}}{v^{2}\left(4m^{2}-m_{h}^{2}\right)^{2}}v_{rel}^{2}
\label{annh}
\end{equation}
where $m_{b}$ is the mass of the $b$ quark (the process with
$b\overline{b}$ in the final state is dominant, since
$\sigma_{h}\propto m_{f}^{2}$).
\subsection{Thermally averaged cross sections}
To compute the thermally averaged cross section it is useful to
observe that in both cases
\begin{equation}
\sigma_{j}\,v_{rel}=b_{j}\,v_{rel}^{2}
\end{equation}
where $j=Z,h$, respectively for $Z$ exchange and for $h$ exchange.
Performing the thermal average we obtain
\begin{equation}
<\sigma_{j} v_{rel}>=6\frac{b_{j}}{x}
\end{equation}
so in our case the value of $\sigma_{0}$ is given by
\begin{equation}
\sigma_{0}=6\left(b_{Z}+b_{h}\right)
\end{equation}

\section{LNP-nucleus elastic cross sections}\label{DirDetApp}
\subsection{Spin-independent cross section}
The elastic scattering process is: $\chi_{l}\;\mathcal{N}\rightarrow
Z^{*}\rightarrow \chi_{l}\;\mathcal{N}$\newline where $\mathcal{N}$
is a generic nucleus. The vertex between quarks and the Higgs boson
is given by the Standard Model Lagrangian and results in
\begin{equation}
\frac{g}{c_{W}}Z^{\mu}J^{0}_{\mu}
\end{equation}
The weak neutral current of the quarks $J^{0}_{\mu}$ is of the form
$\sum_{q}\overline{q}\gamma^{\mu}\left(c_{V}^{q}-c_{A}^{q}\gamma^{5}\right)q$,
where the parameters $c_{V}^{q}$ e $c_{A}^{q}$ are known as function
of the Weinberg angle only. We have two different contributions to
the amplitude: the quarks vector current and the quarks axial-vector
current. We can neglect the first contribute for nonrelativistic
LNP. The vector-axial contribution is described by the effective
lagrangian
\begin{equation}
\mathcal{L}_{axial}=\left(V_{1l}V_{2l}\right)\overline{\Psi_{L}}\gamma^{\mu}\gamma^{5}\Psi_{L}\;
\sum_{q}\zeta_{q}\,\overline{q}\gamma_{\mu}\gamma^{5}q
\end{equation}
where we define $\zeta_{q}\equiv 2\sqrt{2}\,G_{F}c_{A}^{q}$.

We introduce the parameters
\begin{equation}
\begin{array}{l}
a_{p}=\sum_{q}\frac{\zeta_{q}}{\sqrt{2}G_{F}}\Delta q^{(p)}=\sum_{q}2c_{A}^{q}\Delta q^{(p)}\\
a_{n}=\sum_{q}\frac{\zeta_{q}}{\sqrt{2}G_{F}}\Delta
q^{(n)}=\sum_{q}2c_{A}^{q}\Delta q^{(n)}
\end{array}
\end{equation}
\begin{equation}
\Lambda=\frac{a_{p}<S_{p}>+a_{n}<S_{n}>}{J}
\end{equation}
The quantity $J$ is the total angular momentum of the nucleus,
$<S_{p}>$ is the expectation value of the spin content of the proton
group in the nucleus, and similarly for $<S_{n}>$. The total cross
section is \cite{Jungman:1995df}
\begin{equation}
\sigma_{Z}\left(LN\rightarrow
LN\right)=\frac{32}{\pi}\left(V_{1l}V_{2l}\right)^{2}G_{F}^{2}m_{r}^{2}\Lambda^{2}J(J+1)
\label{formulaspinDEP}
\end{equation}
where $m_{r}$ is the reduced mass of the system LNP-nucleus

\subsection{Spin-independent cross section}
The elastic scattering process is:
$\chi_{l}\;\mathcal{N}\rightarrow
h^{*}\rightarrow\chi_{l}\;\mathcal{N}$
\newline
The nucleonic matrix element can be parameterized by
\cite{Barbieri:1988zs}
\begin{equation}
<N|\sum_{q}m_{q}\overline{q}q|N>=f\,m_{N}<N|N>
\;\;\;\;\;\;\;\;\;\;\;\;\; f\simeq 0.3
\end{equation}
and finally the spin-independent cross section results in
\begin{equation}
\sigma_{h}\left(LN\rightarrow LN\right)= \frac{2\xi^{2}f^{2}}{\pi}
\frac{m_{r}^{2}m_{N}^{2}}{m_{h}^{4}v^{2}} \label{formulaspinINDEP}
\end{equation}
where $m_{r}$ is the reduced mass of the system LNP-nucleus.

\section{ElectroWeak Precision Test}\label{EWPTappendix}
As well known, new physics effects to the EWPT are conveniently
represented by the parameters $T$ and $S$, defined by
\begin{equation}
T=\frac{\Pi_{33}(0)-\Pi_{WW}(0)}{\alpha_{em} \;m_{W}^{2}}
\;\;\;\;\;\;\;\;S=\frac{4s_{W}c_{W}}{\alpha_{em}} \Pi^{'}_{30}(0)
\label{TS}
\end{equation}
in terms of the vacuum polarization amplitudes
\begin{equation}
i \Pi_{ij}^{\mu\nu}(q)=\eta^{\mu\nu} \Pi_{ij}\left(q^{2}\right) +
\;\; q^{\mu}q^{\nu} \;\;terms
\end{equation}
with $i,j=3,0,W$ for $W_{3}^{\mu}$, $B^{\mu}$, $W^{\mu}$
respectively.

Expressions for the vacuum polarization amplitudes produced by
fermions coupled to a generic gauge boson are known
\cite{Barbieri:2006bg}. For a fermion loop with internal masses $m_{1}$
and $m_{2}$ and a vector coupling
$V_{\mu}\overline{\Psi}_{1}\gamma^{\mu}\Psi_{2}$ it is
\begin{eqnarray}
\Pi(0) & = & \frac{1}{16\pi^{2}}\left[
\left(m_{1}-m_{2}\right)^{2}\ln\frac{\Lambda^{4}}{m_{1}^{2}m_{2}^{2}}-2m_{1}m_{2}\right.
{}\nonumber\\ & &
{}+\left.\frac{2m_{1}m_{2}\left(m_{1}^{2}+m_{2}^{2}\right)-m_{1}^{4}-m_{2}^{4}}{m_{1}^{2}-m_{2}^{2}}
\ln\frac{m_{1}^{2}}{m_{2}^{2}}\right]
\end{eqnarray}

\begin{eqnarray}
\Pi^{'}(0) & = & \frac{1}{24\pi^{2}}\left[
-\ln\frac{\Lambda^{4}}{m_{1}^{2}m_{2}^{2}}-
\frac{m_{1}m_{2}\left(3m_{1}^{2}-4m_{1}m_{2}+3m_{2}^{2}\right)}{\left(m_{1}^{2}-m_{2}^{2}\right)^{2}}\right.
{}\nonumber\\ & &
{}+\left.\frac{m_{1}^{6}+m_{2}^{6}-3m_{1}^{2}m_{2}^{2}\left(m_{1}^{2}+m_{2}^{2}\right)+6m_{1}^{3}m_{2}^{3}}
{\left(m_{1}^{2}-m_{2}^{2}\right)^{3}}\ln\frac{m_{1}^{2}}{m_{2}^{2}}\right]
\end{eqnarray}
For an axial coupling the results are obtained by letting
$m_{1}\rightarrow-m_{1}$ in the previous expressions. These results
are valid for Dirac fermions, for Majorana fermions there is an
extra factor of 2. $\Lambda$ is a cutoff of the loop integral which
disappears in the overall expressions (\ref{TS}).

We define for convenience
\begin{equation}
\widetilde{A}\left(m_{1},m_{2}\right)\equiv
\frac{1}{2\alpha_{em}v^{2}}\Pi\left(0\right)
\end{equation}

\begin{equation}
\widetilde{F}\left(m_{1},m_{2}\right)\equiv 4\pi
\Pi^{'}\left(0\right)
\end{equation}

\bibliography{dmandhiggs}

\end{document}